\documentclass[journal=ancac3,manuscript=article]{achemso}

\usepackage{chemformula} 
\usepackage[utf8]{inputenc} 
\usepackage[T1]{fontenc} 
\usepackage{graphicx} 
\usepackage{wrapfig}
\usepackage{lineno}

\author{Adrian Hemmi}
\affiliation{Physik-Institut, Universit\"at Z\"urich, 8057 Z\"urich, Switzerland}

\author{Ari Paavo Seitsonen}
\affiliation{D\'{e}partement de Chimie, \'{E}cole Normale Sup\'{e}rieure, F-75005 Paris, France}

\author{Thomas Greber}
\affiliation{Physik-Institut, Universit\"at Z\"urich, 8057 Z\"urich, Switzerland}

\author{Huanyao Cun}
\email{hycun1@physik.uzh.ch}
\affiliation{Physik-Institut, Universit\"at Z\"urich, 8057 Z\"urich, Switzerland}
\email{hycun1@physik.uzh.ch}

\title{The Winner Takes It All: Carbon Supersedes Hexagonal Boron Nitride with Graphene on Transition Metals at High Temperatures}


\begin{document}


\begin{tocentry}
 \includegraphics[scale=0.36]{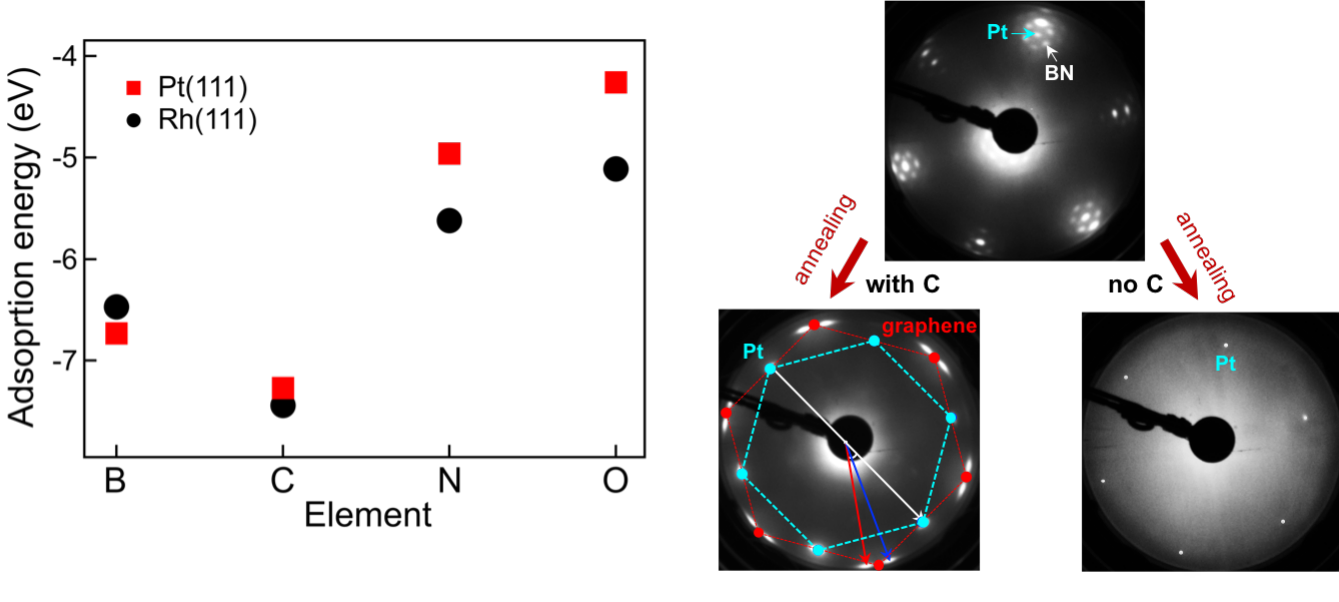}
  \label{fig:TOC}
\end{tocentry}









\begin{abstract}

The production of high-quality hexagonal boron nitride ($h$-BN) is essential for the ultimate performance of two-dimensional (2D) materials-based devices, since it is the key 2D encapsulation material. Here, a decisive guideline is reported for fabricating high-quality $h$-BN on transition metals:  It is crucial to exclude carbon from $h$-BN related process. Otherwise carbon prevails over boron and nitrogen due to its larger binding energy, thereupon forming graphene on metals after high-temperature annealing. We demonstrate the surface reaction-assisted conversion from $h$-BN to graphene with high-temperature treatments. The pyrolysis temperature T$_{p}$ is an important quality indicator for $h$-BN/metals. When the temperature is lower than T$_{p}$, the quality of $h$-BN layer is improved upon annealing. While the annealing temperature is above T$_{p}$, in case of carbon-free conditions, the $h$-BN disintegrates and nitrogen desorbs from the surface more easily than boron, eventually leading to clean metal surfaces. However, once the $h$-BN layer is exposed to carbon, graphene forms on Pt(111) in the high-temperature regime. This not only provides an indispensable principle ({\it{avoid carbon}}) for fabricating high-quality $h$-BN materials on transition metals, but also offers a straightforward method for the surface reaction-assisted conversion from $h$-BN to graphene on Pt(111).

\end{abstract}


\section{Introduction}

Two-dimensional (2D) materials \cite{Novoselov2004} as an emerging platform have been widely explored for ultimate performance of heterostructures and devices due to their superior electronic, mechanical, chemical and thermal properties \cite{Geim2013}. These applications range from sensors and mechanical detectors \cite{Cartamil-Bueno2017} to ultimately thin membranes \cite{Surwade2015}, nanophotonics \cite{Tran2015} and in particular electronics beyond the silicon technology \cite{Banszerus2015}. It is well-known that 2D materials are delicate and require to be well encapsulated with an insulator that does not degrade their performance. On the other hand, the 2D materials needs to be large scale with sufficiently high quality, in particular the 2D encapsulation material. Hexagonal boron nitride ($h$-BN) is the key 2D packaging material since it is flat, transparent, impermeable, chemically inert, electronically insulating yet thermally conductive \cite{Dean2010}. 

However, it is a big challenge to fabricate large-scale $h$-BN of ultimately high quality. Currently, the best-known method to produce scalable high-quality $h$-BN is chemical vapor deposition (CVD) in ultra-high vacuum (UHV) environment with precursors containing boron and nitrogen on transition metals \cite{Nagashima1995,Corso2004,Auwarter2019,Chen2020,Hemmi2021}. High-quality $h$-BN is achieved by optimization of the preparation parameters, for instance the precursor pressure \cite{Cavar2008} or the temperature \cite{Hemmi2021}. For Ir(111) an upper temperature limit to prepare boron nitride is reported, and a boron-based borophene layer is formed at growth temperature of 1100$^\circ$C \cite{Omambac2021}. In addition, post-deposition processing can improve the quality, e.g., in terms of mosaicity and strain variations, if the CVD-grown boron nitride layer is back-transferred on a clean and crystalline metal and followed by subsequent annealing \cite{Cun2020}.

While the CVD synthesis of $h$-BN on transition metals has been well studied, the post-CVD structural changes with high-temperature (HT) annealing or the influence of a post-CVD exposure to contaminations like carbon, on the other hand, have rarely been addressed. Previously, 2D hexagonal boron–carbon–nitrogen layers have been synthesized on metals intentionally via various techniques \cite{Beniwal2017,Khanaki2017,Cheng2019}. For the transition metals like Ni, Co\cite{Tian2018,He2019} and Fe\cite{Babenko2020} that used for segregation growth of $h$-BN, carburization was applied to tune the nucleation density and BN layer thickness at growth temperature. In the case of $h$-BN/Cu, in-plane $h$-BN-graphene structures form when the temperature is above 900$^\circ$C, while the stacked graphene/$h$-BN were observed when the temperature is below 900$^\circ$C \cite{Angizi2020}. In the case of Ir(111), nitrogen completely disappears from the surface \cite{Omambac2021}. Although previous reports on $h$-BN fabrication with processes where carbon species have been intentionally added\cite{Khanaki2017,Tian2018,He2019,Babenko2020}, the fact that the best quality BN is obtained at growth temperatures near the dissociation of the BN\cite{Hemmi2021} imposes that carbon, if present will be integrated in the structures and thus alter the intrinsic properties of $h$-BN.

Here, the structural changes of $h$-BN on Pt(111) after CVD with HT treatments are studied systematically, and the temperature limits of stable $h$-BN layers are identified. We demonstrate that even a tiny carbon exposure ($< 10^{-9}$ mbar) can substitute boron and nitrogen on metal surfaces at high temperatures, which differs from the previous work by Kim et al., where CH$_4$ was intentionally dosed as carbon source to form in-plane graphene/h-BN heterostructures\cite{Kim2015}. In the HT regime, pyrolysis of $h$-BN occurs, i.e., the boron-nitrogen bonds are broken. That is, in the presence of carbon, annealing $h$-BN above the pyrolysis temperature results in the formation of graphene on Pt(111). Thus, it is crucial to exclude carbon during $h$-BN processes to attain a high-quality material. Otherwise, carbon prevails compared to boron and nitrogen due to a larger binding energy of carbon on metals. This provides the decisive requirement during the $h$-BN processing (no carbon contamination) and paves the way to scalable and high-quality $h$-BN fabrication on metals. Furthermore, it also provides a straightforward method for the surface reaction-assisted conversion from $h$-BN to graphene.

\section{Results and Discussions}

\begin{figure}[H]
\begin{center}
\includegraphics[scale=0.75]{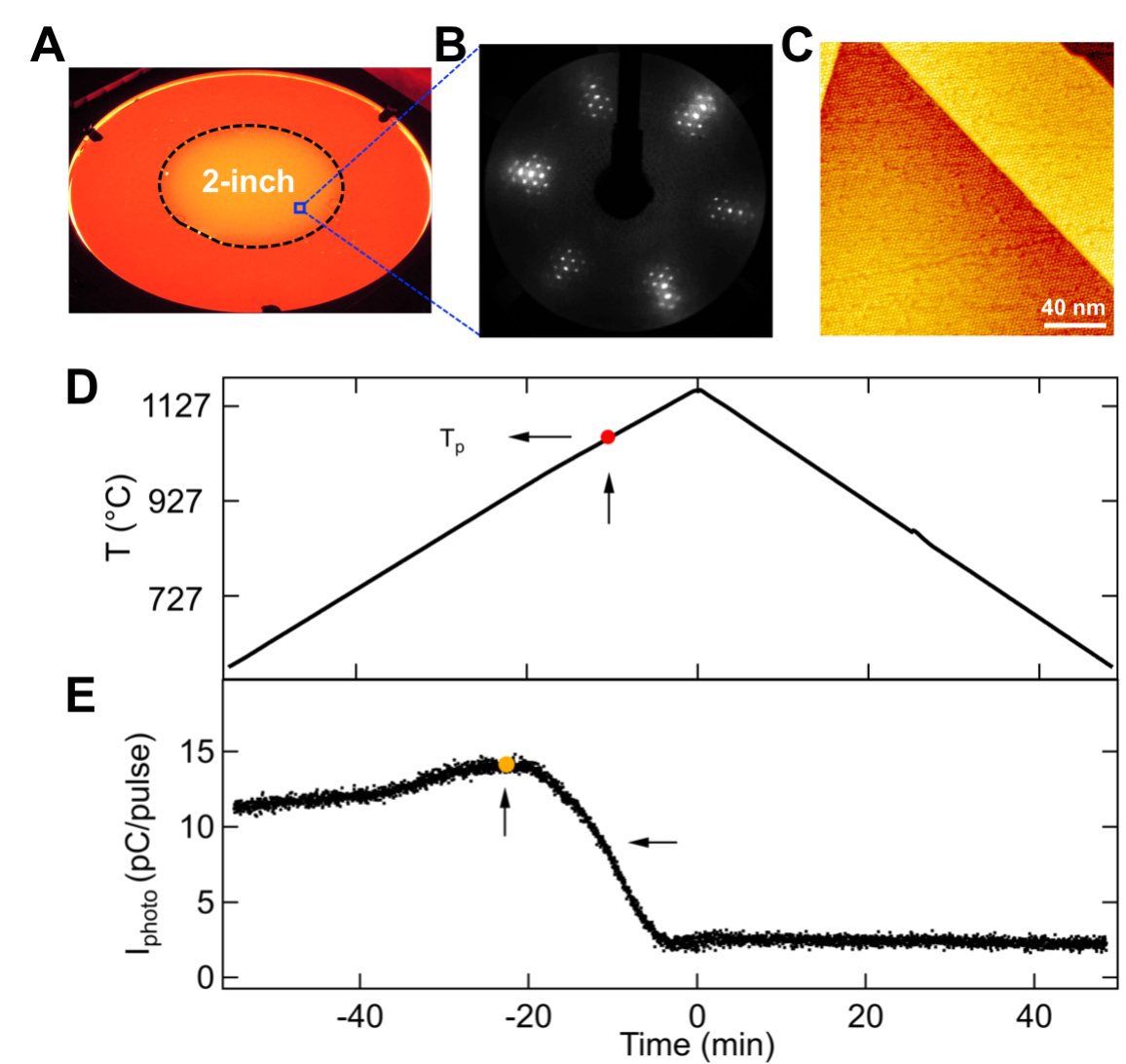}
\caption{{\textbf{\small CVD growth and pyrolysis of $h$-BN on Pt(111).}}
(A) Photo of a 2-inch Pt(111) thin film wafer at high temperature during CVD growth. The dashed circle represents the 2-inch size of the wafer. (B) LEED pattern of $h$-BN/Pt(111) with electron energy of 100 eV acquired with in-situ scanning xy-LEED. (C) Room temperature STM (190~$\times$~190~nm$^2$) image of pristine $h$-BN monolayer on Pt(111). U = -1.20~V, I = 0.50~nA. (D,E) Pyrolysis of $h$-BN on Pt(111) grown at 932$^\circ$C. (D) Temperature as a function of time. The heating rate is 10$^\circ$C/min, t=0 is set to the maximum temperature of 1160$^\circ$C. (E) The photo-current shows the restoration of a high work function surface upon pyrolysis. The orange dot indicates the time when the temperature at which the $h$-BN layer was grown (932$^\circ$C) is reached. The horizontal arrow points to half of the total change in yield at 1057$^\circ$C, which defines the pyrolysis temperature T$_p$ (corresponding to the red dot in (D)).
}
\label{F1}
\end{center}
\end{figure}

The model systems in this report are $h$-BN monolayers on single crystalline rhodium \cite{Corso2004,Berner2007} at 4-inch wafer scale \cite{Gsell2009,Hemmi2014} and 2-inch platinum wafers. Single layer $h$-BN on Pt(111) is fabricated with UHV-CVD with borazine as precursor \cite{Hemmi2021}. Figure \ref{F1} depicts CVD-grown $h$-BN on single crystalline Pt(111) and the pyrolysis protocol. 

Figure \ref{F1}A is a picture of a 2-inch wafer during the fabrication process at high temperature. The 2-inch Pt(111) thin film wafer -- marked with a dashed circle -- is mounted on a 4-inch wafer sample holder. After CVD growth, the wafer is characterized with scanning low-energy electron diffraction (xy-LEED) in the same UHV instrument (UHV$^{\#}$1), and one representative LEED pattern (E=100 eV) is displayed in Figure \ref{F1}B, where the clear and sharp hexagonal spots indicate the high degree of single crystallinity of $h$-BN. Such surface morphology is confirmed with large-area scanning tunneling microscopy (STM) at room temperature (Figure \ref{F1}C). 

\begin{figure}[H]
\begin{center}
\includegraphics[scale=0.60]{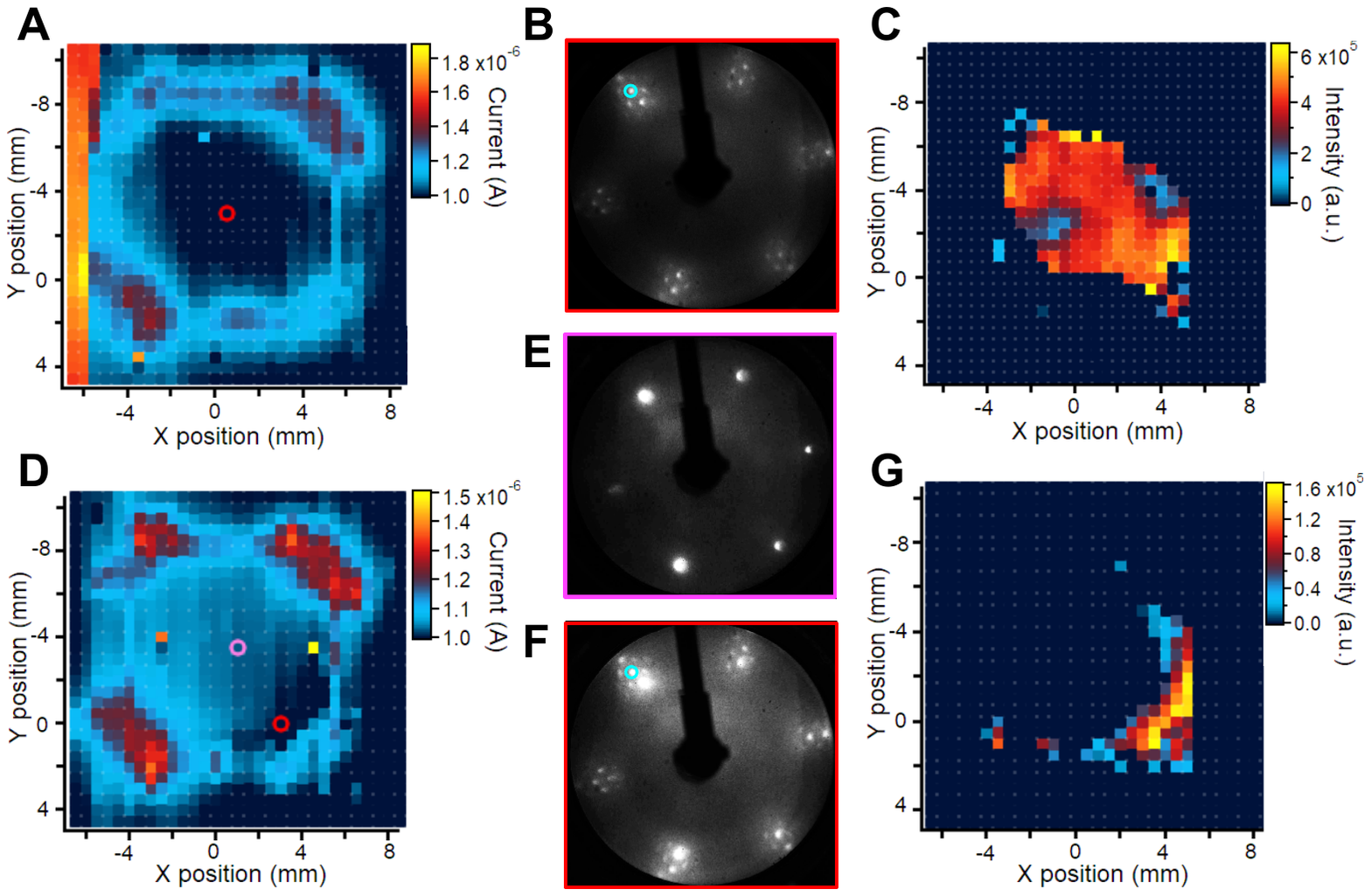}
\caption{{\textbf{\small $h$-BN coverage on Pt(111) before and after annealing to the pyrolysis temperature.}}
(A,D) xy-LEED $h$-BN/Pt(111) sample current map before (A) and after (D) annealing to 1200$^\circ$C. (B,E\&F) LEED patterns (E = 100 eV) of the spots marked in (A) and (D), where (B) is the red circle in (A), while (E,F) are the magenta (E) and red (F) circles in (D). (C,G) are spatial distributions of the marked $h$-BN LEED peaks with light-blue circles in (B) and (F). After annealing to the pyrolysis temperature, the LEED pattern of clean Pt(111) appears on nearly the entire sample, except for the bottom-right corner observed with a small area of intact $h$-BN LEED pattern, which mostly attributes to the sample holder clip.
}
\label{F5}
\end{center}
\end{figure}

Temperature stability of 2D materials is an essential property that can, to some extent, be tuned, though there is a maximum temperature \cite{Omambac2021}. Our data indicate that the thermal BN disintegration moves across the surface similar to the etching of $h$-BN in the presence of oxygen \cite{Goriachko2008} (see partial BN removal in Figure \ref{F5}G). Figure \ref{F1}D shows the temperature treatment for the pyrolysis of a $h$-BN layer on Pt(111) grown at 932$^\circ$C. The wafer is heated up with a constant temperature rate of 10$^\circ$C/min to 1160$^\circ$C and then cooled down with -12$^\circ$C/min. The middle and bottom panels in Figure S1 show the mass spectrometer signals of m/q=11 and m/q=14, where the corresponding partial pressure p$_{11}$ and p$_{14}$ are inferred from the integral of the mass spectra and the total pressure readings. The asymmetry of the p$_{14}$ peak with respect to the temperature indicates the pyrolysis that leads to a desorption of nitrogen into the gas phase. 

The symmetric part of the signal stems from the BN heater. The m/q=11 peak that corresponds to the most abundant $^{11}$B isotope is more than two orders of magnitude weaker. This is understood by dissolution of boron into the Pt and its back-segregation to the surface upon cooling.
During annealing, the $h$-BN/Pt(111) surface gets cleaner. The pyrolysis ignites at the defects and leads to a desorption of nitrogen more easily than boron into the gas phase. The process and the state of the surface are directly monitored with an in-situ photoemission yield setup \cite{Hemmi20142} (Figure \ref{F1}E). The assignment that boron resides on the substrate and that nitrogen desorbs more easily is supported by a subsequent x-ray photoelectron spectroscopy (XPS) elemental analysis and by density functional theory (DFT) calculations below \cite{Hohenberg_1964_PR}.

\begin{figure}[H]
\begin{center}
\includegraphics[scale=0.75]{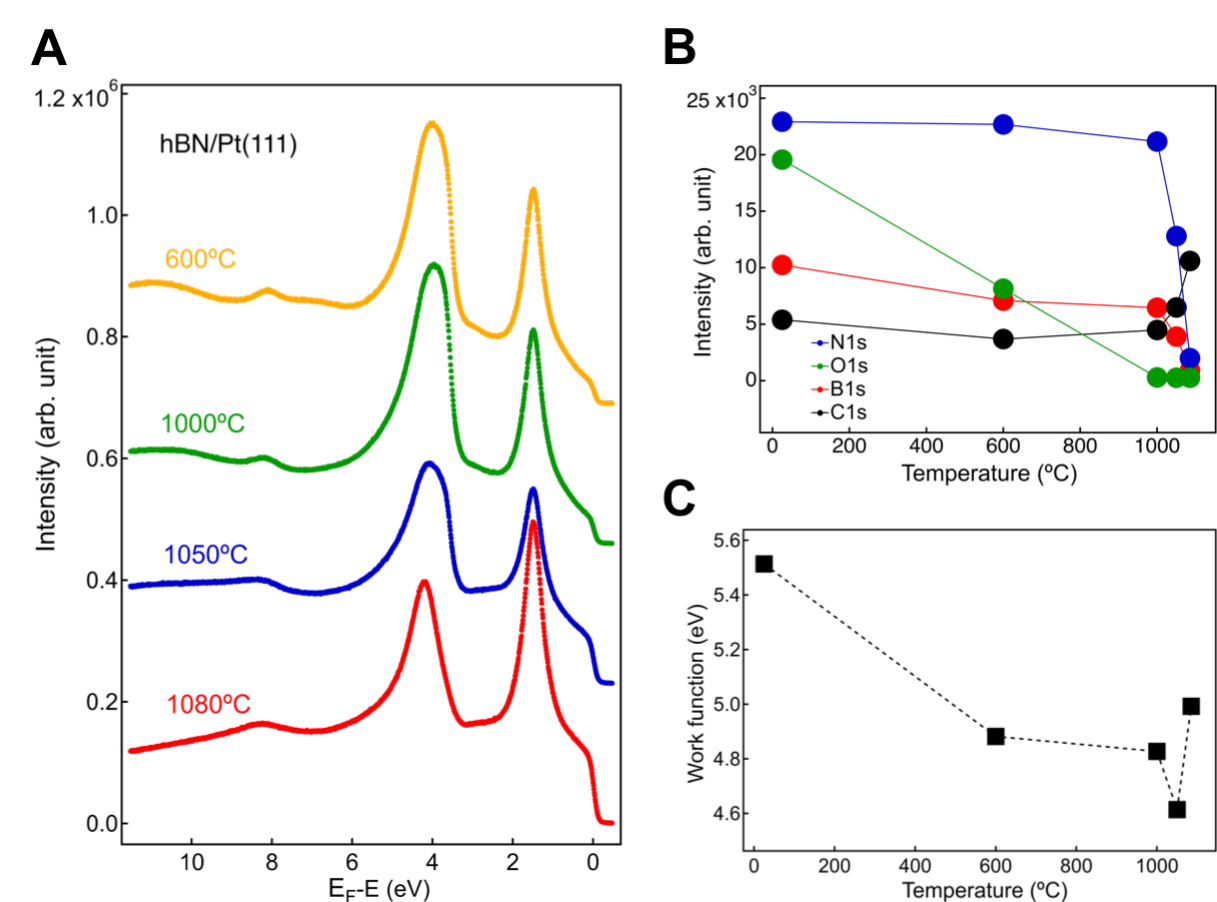}
\caption{{\textbf{\small Structural and electronic properties of $h$-BN/Pt(111) upon annealing treatments.}}
(A) Annealing temperature dependent UPS (He I$_\alpha$, $\hbar\omega = 21.2$ eV) spectra of $h$-BN/Pt(111) after annealing to 600$^\circ$C (orange), 1000$^\circ$C (green), 1050$^\circ$C (blue) and 1080$^\circ$C (red), respectively. The work function changes of \mbox{$h$-BN}/Rh(111) is reflected in modifications of the $\sigma$ band. (B) Integrated intensities of Mg K$\alpha$ XPS ($\hbar\omega = 1253.6$ eV) B1s (red), N1s (blue), C1s (black) and O1s (green) after annealing to different temperatures. Clearly intensities of B1s, N1s and O1s decrease while intensity of C1s increases. (C) Work function as a function of the annealing temperatures. The work function raises once the temperature reaches the pyrolysis temperature.
}
\label{F3}
\end{center}
\end{figure}

Figure \ref{F1}E shows the photoelectron yield as a function of time. Below the growth temperature the increase of the yield is assigned to the higher photoemission probability at higher temperatures \cite{Hechenblaikner2012}. When the temperature is above 932$^\circ$C, pyrolysis sets in as can be seen from the decreasing yield. The photo-current indicates the restoration of a surface with a larger work function upon pyrolysis. This fits well with the work function changes due to varying annealing temperatures(Figures \ref{F3}C and S8C). The pyrolysis temperature T$_p$ is defined as the temperature at which the total change in yield is reduced to 50\% for a heating rate of 10$^\circ$C/min. Thus, a T$_p$ of 1057$^\circ$C is found. From this we conclude that the $h$-BN layer was grown merely at the onset of the T$_p$, where the borazine pressure provides abundant boron and nitrogen and prevents $h$-BN decomposition into a boron-rich layer \cite{Omambac2021}. In this context it is important to quote the heating rate and the gas phase composition since pyrolysis is a kinetic process. Furthermore, it also depends on integrity of the layer and catalytic property of the substrate. For $h$-BN/Ru(0001), the decomposition of $h$-BN layer in 10$^{-8}$ mbar O$_2$ atmosphere occurs at temperatures below 777$^\circ$C \cite{Goriachko2008}. 

\begin{figure}[H]
\begin{center}
\includegraphics[scale=0.59]{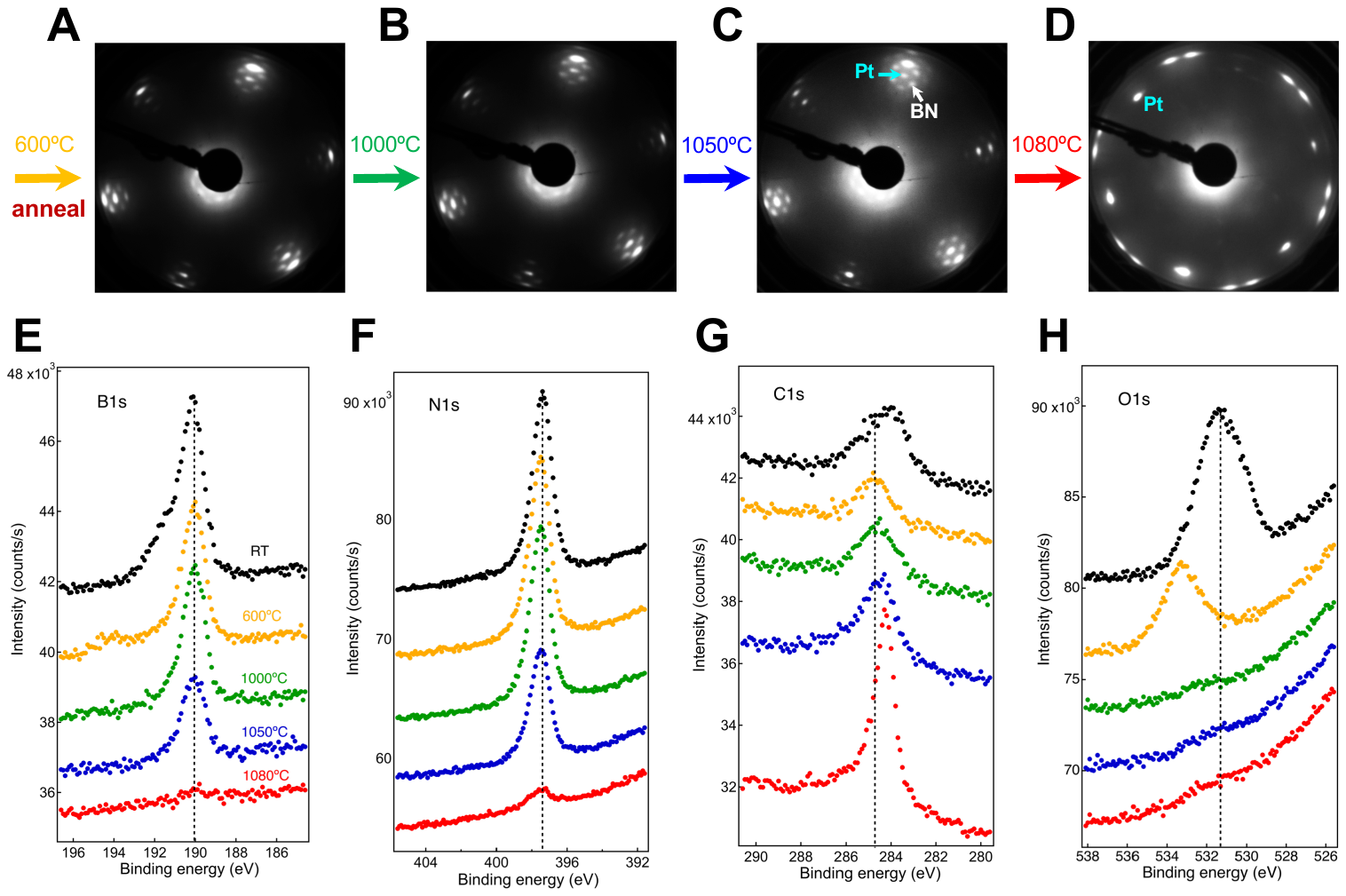}
\caption{{\textbf{\small Structural analysis of high-temperature annealing effects for $h$-BN/Pt(111).}}
(A-D) Temperature dependent LEED patterns for $h$-BN/Pt(111) \mbox{(E=60 eV)}. LEED pattern measured at room temperature after annealing to 600$^\circ$C (A), 1000$^\circ$C (B), 1050$^\circ$C (C) and 1080$^\circ$C (D). Pt spots are marked with light-blue arrow and $h$-BN spots with white arrow. (E-H) Temperature dependent Mg K$\alpha$ XPS ($\hbar\omega = 1253.6$ eV) spectra for B1s (E), N1s (F), C1s (G) and O1s (H). The annealing temperatures are indicated with corresponding colors, including room temperature (black), 600$^\circ$C (orange), 1000$^\circ$C (green), 1050$^\circ$C (blue) and 1080$^\circ$C (red), respectively.
}
\label{F2}
\end{center}
\end{figure}

In order to characterize the pyrolysis process, a series of HT annealing treatments from 600$^\circ$C to 1080$^\circ$C for $h$-BN/Pt(111) and from 600$^\circ$C to 1150$^\circ$C for $h$-BN/Rh(111) were conducted. LEED, XPS  and ultraviolet photoelectron spectroscopy (UPS) measurements were carried out after each annealing step in the same UHV instrument (UHV$^{\#}$2). A representative series of experiments for $h$-BN/Pt(111) are shown in Figure \ref{F2}. In contrast to Figure \ref{F1}, the data in Figure \ref{F2} are recorded after samples transfer in air from CVD system UHV$^{\#}$1 to another analysis chamber UHV$^{\#}$2. After annealing to 600$^\circ$C, the core level peak of B1s has a  chemical shift towards lower binding energy by 0.6 eV, while N1s, C1s and O1s shift towards the higher by 0.1, 0.4 and 2.0 eV, respectively (Figures \ref{F2}E-\ref{F2}H). 

This shows similar B1s peak shift for $h$-BN on Mo(110) \cite{Allan2007}, but differs from $h$-BN on Ir(111), where the double peak of B1s was caused by modulating the growth temperature \cite{Omambac2021}. Although both systems, $h$-BN/Ir(111) and $h$-BN/Mo(110), show that higher growth temperature leads to boron enrichment, the chemical shift directions are not alike \cite{Omambac2021,Allan2007}.
In the present pyrolysis experiments, the O1s intensity decreases significantly upon annealing (Figure \ref{F2}H). Notice that the $h$-BN and Pt spots in the LEED patterns get clearer and sharper while the temperature reaches 1000$^\circ$C, indicating the quality improvement of the $h$-BN layer (Figure \ref{F2}B). This is also reflected on the smaller work function when the temperature is below 1000$^\circ$C (Figure \ref{F3}C). However, the intensities of B1s and N1s decrease when temperatures are above 1000$^\circ$C, as shown in Figures \ref{F2}E\&\ref{F2}F. This critical transition temperature is the pyrolysis temperature T$_{p}$. When the temperature is higher than T$_{p}$, the $h$-BN decomposes and nitrogen desorbs from the surface more easily than boron.
By increasing the annealing temperature to 1080$^\circ$C, the atomic ratios change from 0.13 to 0.04 for boron and from 0.13 to 0.02 for nitrogen, and the B/N atomic ratio alters from 1:1 to 2:1. Meanwhile, a significant C1s peak appears (Figure \ref{F2}G), and the $h$-BN LEED spots disappear while new bright sports are visible and coexist with the Pt spots. Due to the notable amount of carbon on the surface after annealing, these special spots are associated with carbon naturally. The detailed analysis will be discussed later.

Figure \ref{F3}A displays the corresponding UPS spectra. The dominant Pt peak near the Fermi energy does not show significant changes while the peak width and height of the band at higher energy is modified upon annealing. This indicates a structure modification of $h$-BN/Pt(111), since it constitutes a BN-band and a Pt5d-band. Accordingly, the work function increases after the $h$-BN decomposition (Figure \ref{F3}C), which indicates enlarged bare Pt(111) surface. To check the structure modifications, the intensities of four main elements 
including B1s, N1s, O1s and C1s are displayed in Figure \ref{F3}B. Clearly, the intensities of B1s, N1s and O1s decrease upon annealing treatments from 600$^\circ$C to 1080$^\circ$C. Further analysis of B/N atomic ratio is shown in Figure S6.  Notice that the B/N ratio is rather stable till the temperature reaches the decomposition threshold (T$_{p}$) of $h$-BN on Pt(111) (black and red) and Rh(111) (blue), where nitrogen desorbs more easily than boron, leading to boron-rich structures on surfaces. This is consistent with the in-situ monitoring photoemission results in Figure \ref{F1}E. The pyrolysis temperature T$_{p}$ is largely related to the growth temperature of $h$-BN layer, and they are positively correlated \cite{Hemmi2021}.

Interestingly, the C1s intensity increases upon annealing treatments (Figures \ref{F2}G\&\ref{F3}B). The atomic ratio of C1s changes from 0.06 after 600$^\circ$C to 0.20 after 1080$^\circ$C annealing. Similar results were obtained for another $h$-BN/Pt sample with smaller annealing steps (Figure S3C). These are interesting yet surprising observations, since the $h$-BN/Pt(111) samples were kept in the electron spectroscopy for chemical analysis (ESCA) chambers for all measurements. The carbon source is identified to be the preparation chamber that has a base pressure of 10$^{-9}$ mbar. Figure S7 shows the C1s intensity changes as function of time. The shown two representative $h$-BN/Pt(111) samples were kept inside of the chamber. Except for the first degas at 600$^\circ$C, it is obvious the longer the samples remain inside of the chamber, the more carbon appears on the surface. Although the carbon contamination in the annealing chamber is rather small ($< 10^{-9}$ mbar), this carbon modifies the surface structure significantly at high temperatures. 

\begin{figure}[H]
\begin{center}
\includegraphics[scale=0.60]{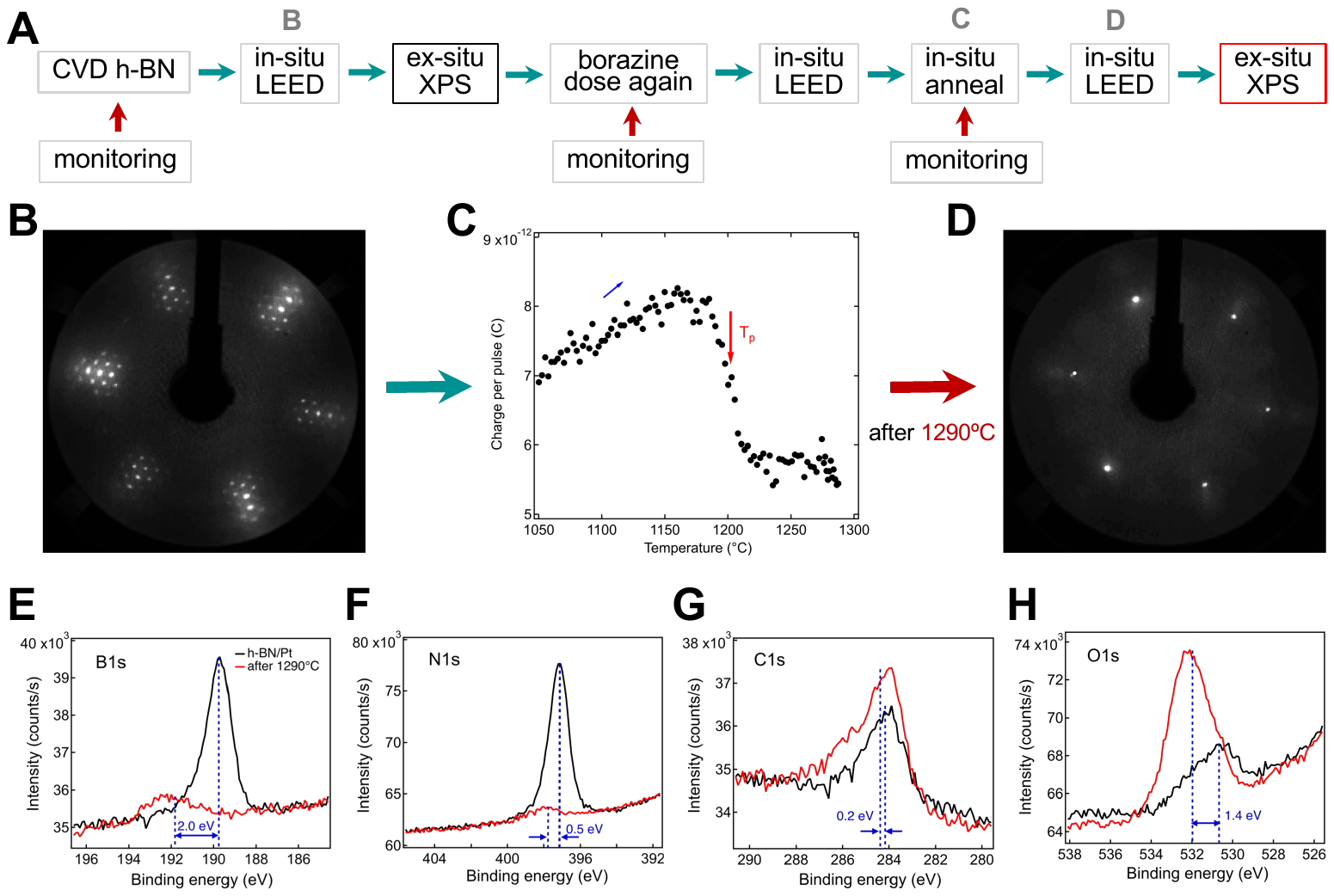}
\caption{{\textbf{\small Structural modification of $h$-BN/Pt(111) excluding carbon contamination before and after high-temperature annealing.}}
(A) The schematic illustrates the experimental work flow, including specific steps where the data in (B,C\&D) are acquired (marked with grey letters). Borazine was dosed again after the first time of ex-situ XPS measurement to exclude carbon contamination. (B) In-situ LEED pattern \mbox{(E=100 eV)} of pristine $h$-BN on Pt(111) prior to the pyrolysis annealing, where clear and sharp $h$-BN and Pt spots are visible. (C) UV flash lamp signal \cite{Hemmi20142} during annealing to 1290$^\circ$C with heating rate of 10$^\circ$C/min. The blue arrow represents the process time direction. A significant signal drop is observed after annealing temperature reaches the pyrolysis temperature T$_p$ of 1200$^\circ$C (red arrow). (D) In-situ LEED structure \mbox{(E=100 eV)} after annealing of $h$-BN/Pt(111) to 1290$^\circ$C, where $h$-BN spots disappear and only Pt spots are visible on the surface. (E-F) Mg K$\alpha$ XPS ($\hbar\omega = 1253.6$ eV) spectra for B1s (E), N1s (F), C1s (G) and O1s (H) before (black) and after (red) annealing to 1290$^\circ$C. The corresponding experimental steps are marked with black and red rectangular in (A). The chemical shifts for B1s (E), N1s (F), C1s (G) and O1s (H) are observed after annealing, and marked with blue dashed lines.
}
\label{F4}
\end{center}
\end{figure}

For comparison, another series of measurements were designed to minimize the influence of carbon. Figure \ref{F4}A exhibits the work flow of the entire process, including steps where the data in Figure \ref{F4}B-\ref{F4}D are acquired (marked with grey letters). It is worth noticing that the $h$-BN/Pt(111) sample was transferred back to the original fabrication chamber of UHV$^{\#}$1 after the first ex-situ XPS in UHV$^{\#}$2, where the sample was degassed thoroughly and borazine was dosed again to ensure a complete $h$-BN layer growth. Such $h$-BN layer involves minimal carbon contamination, and the in-situ xy-LEED displays clear and sharp $h$-BN spots (Figure \ref{F4}B). Figure \ref{F4}C displays the UV flash lamp signal \cite{Hemmi20142} during annealing process to 1290$^\circ$C with a constant temperature rate of 10$^\circ$C/min. The UV lamp signal provides a clear sign when the $h$-BN decomposes and a pyrolysis temperature T$_p$ of 1200$^\circ$C is found (the red arrow in Figure \ref{F4}C). This T$_p$ is correlated to the growth temperature of 1050$^\circ$C, and is higher than the one in Figure \ref{F1}E of 1057$^\circ$C, indicating that T$_p$ can be used as a quality measure for $h$-BN layers. After this HT annealing, the LEED pattern shows only Pt spots and no carbon related spots on the surface. This is a direct comparison with the LEED structure in Figure \ref{F2}D, where the surface was exposed to carbon. 

Figures \ref{F4}E-\ref{F4}H exhibit XPS spectra of B1s, N1s, C1s and O1s before (black) and after (red) HT annealing treatment (the corresponding experimental steps are marked with black and red rectangular in Figure \ref{F4}A). Due to previously mentioned carbon contamination possibility in the preparation chamber of ESCA, the $h$-BN/Pt(111) sample was not degassed after transfer into ESCA, neither before nor after 1290$^\circ$C annealing. That is why a C1s signal is still visible due to short-time (10 min) air exposure for sample transfer (Figure \ref{F4}G). Notice that both B1s and N1s intensities decrease significantly. The B1s intensity declines by -73\% while N1s drops by -91\%. Obviously, the N1s intensity declines more, which is consistent with the in-situ photoemission current measurement in Figure \ref{F1}E and the nitrogen and born partial pressure measurements (Figure S1), which confirms that the adsorption energy of boron and nitrogen differ. The normalized atomic ratios of B/N change from 1:1 to 3:1 upon HT annealing, and lead to boron-rich surface. The slight difference comparing with the B/N ratio changes in Figure \ref{F2} (from 1:1 to 2:1) mainly attributes to the different annealing temperatures (1080$^\circ$C in Figure \ref{F2}), although they both denote boron-rich surfaces. Additionally, both B1s and N1s peaks shift towards higher binding energy, where B1s shifts by 2.0 eV and N1s by 0.5 eV, respectively. After HT annealing, the intensities increase 22\% for C1s and 28\% for O1s. The significant amount of oxygen attributes to the enlarged bare Pt(111) surface, and it is known that bare Pt can be easily oxidized in air. This holds well for xy-LEED results, where only the Pt(111) pattern is visible across the entire sample (Figure \ref{F4}D). It indicates that the remaining boron and nitrogen do not form an ordered structure, thus no $h$-BN LEED spots are observed.

To gain information of the pyrolysis mechanism, a $h$-BN/Pt(111) sample was annealed to T$_p$, and followed by an immediate cooling. Such that a potential pyrolysis process is "frozen" as it was ongoing before the temperature drops. Nevertheless, "freezing" of such process is difficult, because defects in $h$-BN monolayers on transition metals tend to self-heal at high temperatures \cite{Cun20142}. However, remaining boron and nitrogen in Figure \ref{F4}, whose coverage is nearly 18\% monolayer with potential intact BN coverage of $\sim10$\%, do not recover the ordered $h$-BN on Pt(111). 

Figures \ref{F5}A and \ref{F5}D display the sample current maps prior and after annealing to T$_p$. The boron nitride principal reflections -- marked in Figure \ref{F5}B -- are found on the complete sample area in Figure \ref{F5}C prior to the HT treatment. After this T$_p$ annealing process, the majority of the sample only displays Pt(111) patterns (Figure \ref{F5}E), but a small area of $h$-BN is observed at the bottom-right corner (Figures \ref{F5}F and \ref{F5}G). Because the $h$-BN coverage in this area is less than the remaining potential BN coverage in Figure \ref{F4}, it is conjectured that this corner was less affected by the pyrolysis process. This mostly relates to the sample holder clip at the bottom-right corner that leads to a temperature gradient. Therefore, the pyrolysis mechanism of $h$-BN layer on a transition metal appears to be a lateral process as it was observed for $h$-BN on Ru(0001) \cite{Goriachko2008}.

\begin{figure}[H]
\begin{center}
\includegraphics[scale=0.59]{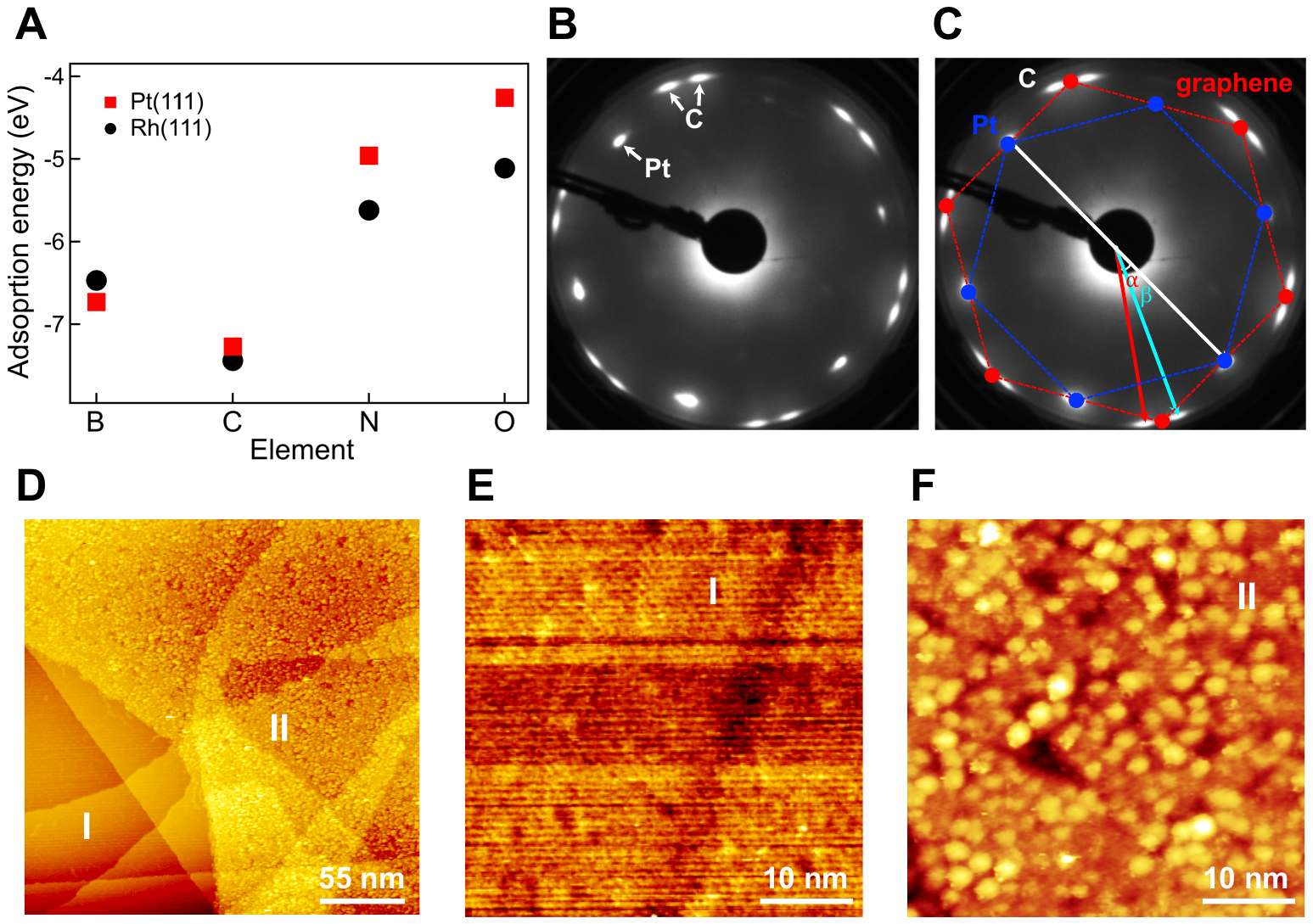}
\caption{{\textbf{\small Adsorption energy differences and graphene formation on transition metal upon annealing.}}
(A) Adsorption energies of boron, carbon, nitrogen and oxygen on Pt(111) and Rh(111). It is obvious that carbon has the largest adsorption energy comparing with the other elements, indicating carbon prevails over boron and nitrogen. (B,C) LEED patterns of $h$-BN/Pt(111) with carbon contaminations after high-temperature annealing treatment without (B) and with (C) embedded graphene (red) and Pt (blue) structure. Besides the Pt spots, the two bright spots are carbon related. Note these spots locate on the same ring of the common graphene phase that is rotated  30$^\circ$ relative to high symmetry Pt direction. The relative angles of these two carbon spots with Pt are marked as $\alpha$ and $\beta$, where $\alpha$ = 35$^\circ$ and $\beta$ = 24$^\circ$, respectively. (D-F) Room temperature STM images of high-temperature annealed $h$-BN/Pt(111). (D) Coexistence of Pt surface and graphene structure. \textbf{I} represents bare Pt(111) surface (E) and \textbf{II} is the carbon-related graphene structure. U = -1.50 V, I = 0.50 nA.
}
\label{F6}
\end{center}
\end{figure}

To better understand the mechanism of the annealing process for $h$-BN on metals with and without carbon, DFT calculations were conducted. The adsorption geometry of the elements B, C, N and O were optimized on Pt(111)-(3$\times$3) and Rh(111)-(3$\times$3) unit cells. The corresponding adsorption energies were calculated and are depicted in Figure \ref{F6}A, where Pt(111) system is in red and the Rh(111) in black. Both Pt and Rh show the same tendency: Comparing with N, B has larger adsorption energy, which explains the boron enrichment after pyrolysis. However, carbon has the largest and oxygen has the smallest adsorption energy. This indicates that oxygen desorbs from the surface prior to the others during an annealing procedure, while carbon is the winner and tends to remain and supersede the other elements. Therefore, the extra spots appeared after HT treatment are carbon related (Figures \ref{F2}D\&\ref{F6}B). The STM image in Figure \ref{F6}D displays the surface morphology of the coexistence of the Pt surface (I) and carbon structures (II), as illustrated by the zoom-in STM images in Figures \ref{F6}E and \ref{F6}F, respectively.

The graphene/Pt(111) system has been studied intensively, and it is well-known that graphene tends to form various rotational domains on Pt(111) due to the very weak interactions between graphene and Pt(111) \cite{Sutter2009,Gao2011,Klimovskikh2014,Rodriguez2015,Cushing2015}. Thus, it is common that the graphene LEED pattern shows the presence of a ring, which indicates different orientations on Pt(111). In LEED patterns, the angle of graphene relative to the Pt spots ranges from 0$^\circ$ to 60$^\circ$ depending on different growth conditions. The most common angles are 19$^\circ$ and 30$^\circ$ \cite{Gao2011,Klimovskikh2014,Rodriguez2015,Cushing2015}. After embedding the graphene 30$^\circ$ spots (red) and Pt (blue) lattice principal diffractions into the special LEED pattern in Figure \ref{F6}B, it is obvious that the two bright carbon related spots locate on a ring with the same polar angle.

The relative angles between these two carbon spots with Pt are $\alpha$ (35$^\circ$) and $\beta$ (24$^\circ$). The angle of 24$^\circ$ is the same as reported by Cushing et al. \cite{Cushing2015}. This reveals that the carbon-related structure on Pt(111) surface is graphene with two main orientations relative to Pt. It is noticed that tuning temperature and annealing time can lead to the coexistence of $h$-BN and Pt (Figure S9B), as well as boron-carbon-nitrogen layers,  even the coexistence of $h$-BN and graphene (Figures S2G\&S10B)\cite{Beniwal2017,Cheng2019,Hemmi2019}. The process is understood as following: Once the decomposition sets in near the pyrolysis temperature T$_{p}$, N and B vacancies are created on the surface. If the sample is exposed to carbon, C atoms can adsorb at the defects and penetrate through the vacancies, thus reach the metal surface and form more stable graphene upon the accumulation of more carbon atoms at high temperature (higher than T$_{p}$).

The DFT results suggest that such graphene formation is also expected on Rh(111). This agrees well with our previous work, where carbon was brought to $h$-BN/Rh(111) externally with an electrochemical process. After annealing to 830$^\circ$C, graphene was formed and embedded as small island in the $h$-BN layer on Rh(111) surface, although the graphene portion was relatively small \cite{Hemmi2019}. The annealing treatment for $h$-BN/Rh in Figure S4 \& S5 indicates no obvious graphene formation on Rh substrate. This mainly due to the following reasons: (1) The $h$-BN layer has very little defects; (2) The carbon exposure is below the threshold of graphene formation at high temperatures (Figure S5C); (3) Appearance of many additional spots in LEED attributes to a damaged Rh thin film after annealing to 1150$^\circ$C\cite{Gsell2009,Hemmi2014}, which makes the observation of graphene spots rather difficult at such conditions.

\section{Conclusions}

In summary, high-temperature annealing treatment is an efficient approach to improve quality of 2D materials, in particular for $h$-BN on transition metals. We report manifest structural modification caused by a combination of HT annealing and carbon exposure for $h$-BN on Pt(111) and Rh(111). When the annealing temperature is lower than the pyrolysis temperature T$_p$, the quality of $h$-BN is improved, whereas when the annealing temperature is above T$_p$, the $h$-BN layer disintegrates and nitrogen desorbs from the surface more easily than boron. The disintegration mechanism likely follows a lateral reaction scheme. The pyrolysis temperature depends on the growth temperature, thus being an important quality indicator for $h$-BN/metal systems. In the case of carbon free, the bare metal surface can be recovered thermally. In the presence of carbon, introduced either internally (inside of UHV chamber) or externally (outside of UHV), graphene forms on metals upon HT treatments due to its larger adsorption energy as compared to boron and nitrogen. Therefore, it is crucial to exclude carbon contamination in $h$-BN related process to achieve and maintain high-quality $h$-BN material. Our results not only provide a decisive guideline ({\it{avoid carbon}}) for fabricating high-quality $h$-BN materials on transition metals for broad applications, but also offer a simple approach for the surface reaction-assisted conversion from $h$-BN to graphene.

\section{Materials and Methods}
{\bf{CVD with in-situ monitoring \& scanning xy-LEED experiments}}:
The \mbox{$h$-BN} monolayers were prepared in an ultra-high vacuum, CVD cold-wall chamber on 2-inch Pt(111) \cite{Hemmi2021} and 4-inch Rh(111) thin film wafers with the procedures and facility described previously \cite{Hemmi2014}. Prior to all {\mbox{$h$-BN}} preparations, the Pt and Rh substrates were cleaned by Ar sputtering, O$_2$ exposure and annealing cycles until sharp Pt(111) and Rh(111) LEED patterns were observed. Subsequently, $h$-BN was prepared at different substrate temperatures above 727$^\circ$C with borazine (HBNH)$_3$ as precursor with a partial pressure of 10$^{-7}$ mbar. The growth procedures were traced with photoelectron yield measurements with a xenon flash lamp \cite{Hemmi20142}, where photoelectron pulses on an area of 1.0 cm$^2$ were detected at a rate of about 0.6~Hz. 
After growth the $h$-BN structures were investigated in the same apparatus with an OCI BDL600-MCP LEED optics where the diffraction patterns were recorded with a CCD camera. In order to quantify the crystal-quality of the $h$-BN on the wafer scale, the sample was scanned in front of the LEED optics on an x-y piezo motor driven stage. The cycle time for moving the x-y stage and taking an image was 1.2 $\pm$ 0.3 s for positioning and CCD exposure.

{\bf{XPS, UPS, LEED \& STM experiments}}:
To carry out XPS and UPS measurements, the 2-inch $h$-BN/Pt(111) and 4-inch $h$-BN/Rh(111) wafers were cut into 10$\times$10 mm$^2$ size, and were transferred into a Vacuum Generators ESCALAB 220 system \cite{Greber1997}, where XPS, UPS and LEED measurements were carried out. The STM experiments were performed at room temperature with an electrochemical etched tungsten tip in a variable-temperature scanning tunneling microscope (Omicron, VT-STM) \cite{Cun2013}. 

{\bf{DFT Simulations}}:
The DFT calculations were performed using the code \texttt{Quantum ESPRESSO} \cite{Giannozzi_2009_JPCM}, applying the generalized gradient approximation of Perdew, Burke and Ernzerhof (PBE) \cite{Perdew_1996_PRL} as the approximation to the exchange-correlation functional. Five layers, of which three lowest ones were kept fixed, of substrate were used to describe the surface in a slab geometry. The first Brillouin zone was sampled with $6\times{}6$ equi-distant k points, and the occupation numbers broadened with the Fermi-Dirac distribution. Further details of the calculations will be published later.




\begin{acknowledgement}

Financial support by the European Commission (European Union's Horizon 2020 research and innovation programme) under the Graphene Flagship Core 3 (No. 881603) is gratefully acknowledged. H-Y. Cun acknowledges the funding support from the Forschungskredit der University Z\"{u}rich (Grant No. FK-21-116).
We thank Mr. Michael Weinl and Dr. Matthias Schreck from Universit\"{a}t Augsburg, Germany for providing Rh metal substrates, as well as Dr. Steven Brems from Imec Leuven, Belgium for providing Pt metal wafers.

\end{acknowledgement}

\begin{suppinfo}


Supporting information on mass spectra changes during $h$-BN pyrolysis on Pt(111), temperature-dependent structural changes of $h$-BN on Pt(111) and Rh(111) with LEED and XPS  based on a series of high-temperature annealing treatments, analysis of atomic ratio of B/N ratio changes with temperatures, UPS and work function modifications with annealing temperatures, different annealing temperature and duration time effect for $h$-BN/Pt(111), and intermediate LEED state are available in Ref. \cite{SI}.

\end{suppinfo}

\bibliography{Pyrolysis_bibliography}


\end{document}